\documentclass[10pt, a4paper]{article}
\usepackage[top=50pt,bottom=50pt,left=48pt,right=48pt]{geometry}
\usepackage[utf8]{inputenc}
\usepackage[usenames,dvipsnames]{color}  
\usepackage{graphicx} 

\usepackage{caption}
\usepackage{float}
\usepackage{titlesec}
\usepackage{amsmath}
\DeclareMathOperator{\Tr}{Tr} 
\usepackage{graphicx}
\usepackage{amssymb}
\usepackage{authblk}
\begin{document}
\date{23 July 2021}
\title{Probing multipartite entanglement, coherence and quantum information preservation under classical Ornstein-Uhlenbeck noise}
\author[1]{Atta Ur Rahman\thanks{Corresponding Author: Attapk@Outlook.com}}
\author[1]{Muhammad Javed}
\author[1]{Arif Ullah}
\affil[1]{Quantum Optics and Quantum Information Research Group, Department of Physics, University of Malakand, Khyber Pakhtunkhwa, Pakistan}
%\title{Probing multipartite entanglement, coherence and quantum information preservation under classical Ornstein-Uhlenbeck noise}
\maketitle
%\author{Atta Ur Rahman}
%\email{Attazaib5711@gmail.com}
%\address{Quantum Optics and Quantum Information Research Group, Department of Physics, University of Malakand, Khyber Pakhtunkhwa, Pakistan}
%\author{Muhammad Javed}
%\author{Arif Ullah}
%\address{Quantum Optics and Quantum Information Research Group, Department of Physics, University of Malakand, Khyber Pakhtunkhwa, Pakistan}

\begin{abstract}
We address entanglement, coherence, and information protection in a system of four non-interacting qubits coupled with different classical environments, namely: common, bipartite, tripartite, and independent environments described by Ornstein-Uhlenbeck (ORU) noise. We show that quantum information preserved by the four qubit state is more dependent on the coherence than the entanglement using time-dependent entanglement witness, purity, and Shannon entropy. We find these two quantum phenomena directly interrelated and highly vulnerable in environments with ORU noise, resulting in the pure exponential decay of a considerable amount. The current Markovian dynamical map, as well as suppression of the fluctuating character of the environments are observed to be entirely attributable to the Gaussian nature of the noise. Furthermore, the increasing number of environments are witnessed to accelerate the amount of decay. Unlike other noises, the current noise parameter's flexible range is readily exploitable, ensuring long enough preserved memory properties. The four-qubit GHZ state, besides having a large information storage potential, stands partially entangled and coherent in common environments for an indefinite duration. Thus, it appeared to be a more promising resource for functional quantum computing than bipartite and tripartite quantum systems. In addition, we derive computational values for each system-environment interaction, which will help quantum practitioners to optimize the related kind of classical environments.
\end{abstract}
\maketitle
\section{Introduction}
Quantum information processing showed substantial growth in various interdisciplinary research fields in recent years, demonstrating its inevitable significance while contributing hugely to the modern technological world \cite{Blais, Xavier, Khalid, Blok}. The quantum information theory replaced the old conventional classical information theory, which has the capability of processing quantum information with physically guaranteed hidden sharing \cite{Moreno}, better simulations \cite{Cirstoiu}, very large and small scale data measurements \cite{Blais-2, Murali}, and extremely fast information processing \cite{Agueny}. As a result, quantum information processing has developed into a well-understood and examined interdisciplinary area that is increasingly become focused on efficient quantum data transmission, the creation of new quantum applications and physical innovation, as well as the control and development of quantum phenomena and quantum devices.\\
Entanglement is one of the significant non-local features of quantum systems that distinguish them from their classical counterparts \cite{Paneru}. It has received great attention in recent years as a key resource for quantum information processing and communication \cite{Zoller}, quantum cryptography \cite{Yin}, dense coding \cite{Laurenza}, teleportation \cite{Yang}, and the exponential acceleration of certain computing tasks \cite{Gabor, Britt, Raychev}. Furthermore, maximally entangled states are typically used for complete and efficient quantum information execution \cite{Rather, Gratsea}. However, these quantum states are oversensitive, and the initial entanglement is diminished, if not entirely lost upon interaction with external environments \cite{Karpat}. Hence, protecting entanglement in different quantum systems within various kinds of environmental influences should be critically analyzed which may play a key role in processing quantum information. Quantum coherence is indeed a criterion for entanglement and other non-local correlations \cite{Kues}. Hence, it is also a vital physical resource in quantum computation and quantum information processing, of equal importance to entanglement. In fact, the existence and preservation of quantum entanglement are guaranteed by the engineering and protection of quantum coherence between the subsystems of a composite quantum state.\\
For the practical deployment of the quantum mechanical protocols, the entangled quantum systems are coupled with external environments. This coupling induces a variety of disorders, increasing entropy and causing coherence and entanglement decay \cite{Goyal}. The coherence decays over time because of these special environmental defects known as noises, while entanglement can end abruptly in a phenomenon known as entanglement sudden death \cite{Cui}. Thus, in quantum information science, environmental noise is an inevitable physical phenomenon. Therefore, characterizing and optimizing the dynamical properties of quantum correlations in open quantum systems is critical for practical quantum information processing \cite{Cai}. Besides, the effective use of these quantum correlations and avoiding the noisy effects in an environment remain fundamental problems that have been extensively discussed since the previous years.\\
Many researchers have investigated the decoherence processes and entanglement dynamics of various quantum systems, environments, and noises, where important results have been achieved \cite{bbb, far}. In bipartite systems, the decohering effects of environments with different memory properties have been extensively studied \cite{Gour, Amosov, Li, Cui}. Many researchers have implemented various measures to characterize and quantify the dynamical properties of quantum entanglement in bipartite and multipartite open quantum systems in recent developments \cite{Haddadi, Haddadi-2, Guo-2, Walborn}. However, it is shown that entanglement measures do not capture all the quantumness in the multipartite quantum states. Certain separable mixed states with disappearing entanglement can have non-vanishing quantum correlations and thus, can implement deterministic quantum computation. We investigate the principle of entanglement in multi-qubit systems, such as a multipartite state made up of four maximally entangled qubits instead of the simpler ones. Such composite quantum states with many subsystems have infinite applications, such as a many-body system in condensed matter physics \cite{GaoX, Khalid-2}. The investigation of entanglement and coherence for such multipartite states is a complicated task, but due to the large information storage capacity, the full characterization of these may be fruitful for the quantum mechanical tasks.\\
Moreover, classical and quantum frameworks can characterize the system-environment interactions that may cause decoherence. When a quantum system interacts with a random or stochastic field, the classical definition of environments is frequently used \cite{Sergi}. On the other hand, when a system interacts with a quantum bath, such as quantum cavities or a collection of periodic oscillators and spin chains, quantum interpretation is used \cite{Tsang}. Besides, for certain system-environment interactions, there is a classical description that is completely similar to the quantum description. Thus, the classical interaction picture of the system-environment model has been successfully used to investigate the dynamics of open quantum systems and the related quantum correlations and coherence.\\
In this work, we analyze dynamics of the entanglement, coherence, and entropy for four qubit maximally entangled Greene-Bergere-Horne-Zeilinger (GHZ) state exposed to classical fluctuating environments. This multi-qubit state coupling with the external environment induces different noises, such as of Gaussian and non-Gaussian nature \cite{Smolin, Wang}. In particular, we focus on a very fundamental type of noise raising due to the random motion of the particles in the diffusion process of information, known as ORU noise \cite{Carmele, Yang-J}. The origin of this noise lies in the Ornstein-Uhlenbeck process, which is a very frequent term in mathematical and physical sciences. The Brownian random motion of the particles of the environments can cause this noise, especially in the regime of dynamics of quantum systems in the essence of diffusion of information \cite{Eliazar}.\\
Besides, we consider four types of system-environment interactions namely: common, bipartite, tripartite, and independent system-environment interactions. In the first case, all four qubits will be coupled to a single classical environment while to the two environments in the second case. In the third situation, two qubits are coupled to separate environments while we couple the rest of the pair to the third environment. In the final case, we couple each qubit to four independent environments. The primary focus is to present a detailed study of the detrimental behaviour of the noise on the dynamics of four qubit systems under the conditions of such system-environment interactions. Here, entanglement witness \cite{Amaro}, purity \cite{Arthur}, and entropy measure \cite{Kak} will investigate the entanglement, coherence, and quantum information preserved by the four qubit system for a finite time of evolution.\\
This paper is assembled as: In Sec.\ref{$QC$ measures}, the entanglement witness, purity, and quantum information estimators are explained. Sec.\ref{The model and dynamics}, illustrates the physical model, application of the noise, and various types of system-environment interactions. Sec.\ref{Results and discussions} explore the findings and the consequent discussion. In Sec.\ref{Conclusion}, we give the conclusive remarks based on the investigation done.
\section{Measures} \label{$QC$ measures}
This section explores the quantifiers used to evaluate multipartite entanglement, coherence, and quantum information preserved by the system against the classical noise.
\subsection{Entanglement witness}\label{dd}
Multipartite entanglement may also be examined using the Entanglement witness (EW). EW is a simple way to distinguish between separable and entangled states. Following this, for separable and entangled states, EW gives non-negative and negative expectation values, respectively. For partial and true entanglement, various types of witnesses have been considered: bipartite, Schmidt number and multipartite witnesses \cite{Sperling, Hulpke, Brandao}. Entanglement through the EW operation for a time evolved density matrix $\rho(t)$ can be computed as \cite{Bourennane}:
\begin{equation}
EW(t)=-\Tr[\mathrm{E}_o \rho(t)].\label{ewo}
\end{equation}
Where $\mathrm{E}_o$ is the entanglement witness for the GHZ state and $\mathrm{E}_o=\frac{1}{2}\textit{I}-\rho_o$ with $\rho_o$, the initial density matrix of the system. Observable of Eq.\eqref{ewo} indicates the appearance of multipartite entanglement experimentally detectable in the system with negative expectation values. However, the absence of entanglement is not guaranteed by positive expectation values of this measure.
\subsection{Purity}
Purity is a simple computable measure of quantum coherence. It evaluates purity and coherence for the time evolution of a quantum system because of system-environment interaction. We can compute this measure as \cite{Arthur, Tsokeng}:
\begin{equation}
P(t)=\Tr[\rho(t)]^2. \label{purity}
\end{equation}
The state of a $d$-dimensions will be mixed at $P(t)=\frac{1}{m}$, while completely pure and coherent at $P(t)=1$. Any value between the upper and lower bound will suggest the corresponding amount of purity.
\subsection{Entropy}
The quantitative measure of disorder or randomness in a system is called entropy. Shannon came up with the concept of entropy as part of his communication theory. Here, we will use the Shanon entropy to compute the information preservation by a quantum system in its dynamical map while interacting with external classical environments. For a time evolved quantum state $\rho(t)$, \cite{Petz}: give the Shannon entropy as:
\begin{equation}
\mathcal{H}(t)=-\sum_{i=1}^n \lambda_i \log_2 \lambda_i.\label{decoherence}
\end{equation}
For an entangled state with no quantum information loss, $\mathcal{H}(t)=0$, while any other output value of this measure will display the corresponding amount of information loss.
\section{The model and dynamics}\label{The model and dynamics}
We consider four non-interacting identical qubits that were initially prepared as a maximally entangled GHZ state interacting with external random fields. Dynamics and preservation of the entanglement, coherence and quantum information by the system here are of our particular interests. The dynamical map of the four qubits is further studied in four different types of system-environment coupling interactions, namely, common (CSE), bipartite (BSE), tripartite (TSE), and independent system environment (ISE) interaction. In the CSE interaction, we couple four qubits to a single classical system, while to two independent environments in the BSE interaction. In the TSE interaction, the system is coupled with three environments and each qubit is individually coupled with four independent environments. Besides, the dynamics is studied under the Gaussian process generating Ornstein-Uhlenbeck. The total stochastic Hamiltonian state for the system of four qubits in the current case is written as \cite{Tchoffo-2}:
\begin{figure}[ht]
\centering
\includegraphics[scale=0.15]{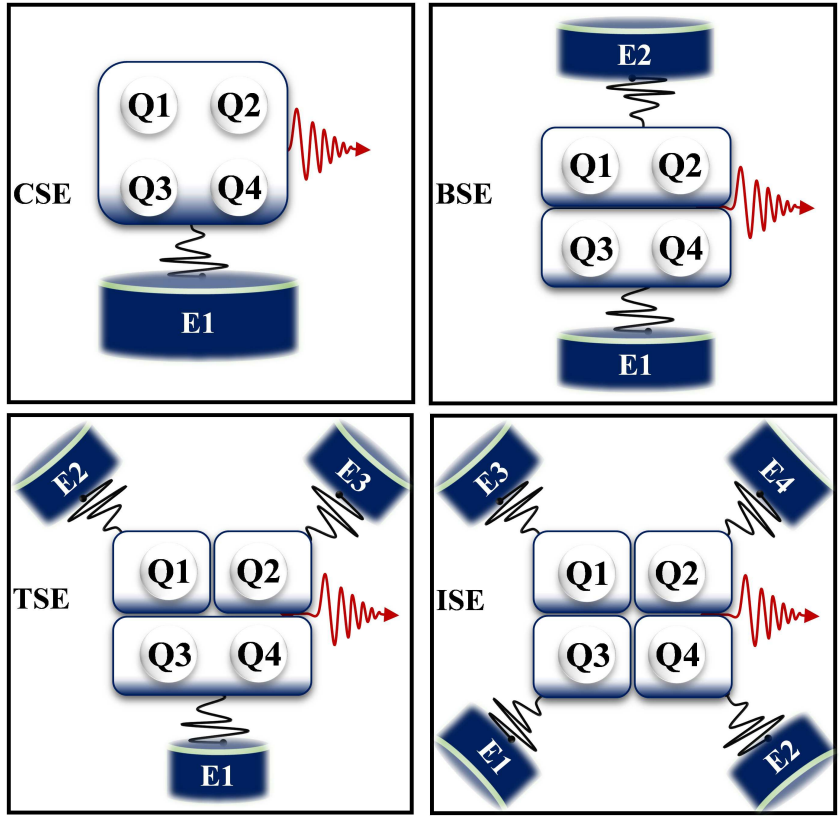}
\caption{Upper panel: Shows the configurations used for four non-interacting maximally entangled qubits, $Q_1$, $Q_2$, $Q_3$ and $Q_4$ coupled to classical environments, $E1$, $E2$, $E3$ and $E4$ shown by the outer boxes in common (left) and bipartite system-environment configuration (right) described by Ornstein-Uhlenbeck noise shown by the blue-greenish shaded regions. The black coloured wavy lines show the system-environment coupling and the reddish wavy lines with reducing amplitude indicates the dynamics and the dephasing effects in the corresponding subspaces of the qubits. The action of the noisy channels over the qubits is represented by the bluish-glow qubit's regions. All the four qubits are interconnected ensuring that the four qubits are prepared in a single state with equal energy splitting represented by the same sizes and shapes of the qubits. Lower Panel: Same as the upper panel but for tripartite (left) and independent system-environment coupling configuration (right).}
\end{figure}
\begin{equation}
H_{ijkl}(t)=H_i(t) \otimes I_j \otimes I_k \otimes I_l+I_i \otimes H_j(t) \otimes I_k\otimes I_l+I_i \otimes I_j \otimes H_k(t)\otimes I_l+I_i \otimes I_j\otimes I_k \otimes H_l(t),\label{hmm}
\end{equation}
where, $H_n(t)$ is the Hamiltonian of the individual subsystem and is defined by $H_n(t)=\mathcal{E}_nI_n+\lambda_{SE}\Delta_n(t)\sigma_n^x$. Here, $\mathcal{E}_n$ is the energy associated with the $n$th qubit, however, in the current case of identical qubits, $\mathcal{E}_i$=$\mathcal{E}_j$=$\mathcal{E}_k$=$\mathcal{E}_l$. The $\lambda_{SE}$ is the coupling constant between the system and environment, while the $\Delta_n(t)$ is the stochastic parameter for the classical environment controlling the fluctuation behavior. The $I_n$ and $\sigma_n^x$ are the identity and Pauli matrices, respectively. Next, the time evolution of the system can be computed as \cite{Tchoffo-2, Rossi}:
\begin{equation}
\rho_{ijkl}(t)=U_{ijkl}(t)\rho_o U_{ijkl}(t)^{\dagger},\label{time evolved density matrix}
\end{equation}
where, $U_{ijkl}(t)$ is the time-unitary operator and is defined as $U_{ijkl}(t)=\exp[\int^{t}_{0}H_{ijkl}(x)dx]$ while we assumed $\hbar=1$. $\rho_o$ is the initial density matrix of the four qubit entangled states and can be written as \cite{Tchoffo-2,Rossi}:
\begin{equation}
\rho_o = [1-p]\frac{I_{(16\times 16)}}{16}+p[\rho_o^{GHZ}],
\end{equation}
where $\rho_o^{GHZ}=\vert GHZ\rangle\langle GHZ\vert$ with $p \in \{0,1 \}$ and $ \vert GHZ\rangle $ is the four qubit maximally entangled Greenberger-Horne-Zeilinger state and is defined as $ \vert GHZ\rangle=\frac{1}{\sqrt{2}}(\vert 0000 \rangle+\vert 1111 \rangle$.
\subsection{Ornstein-Uhlenbeck noise}
To find out detrimental effects of ORU noise, we consider the classical field $\mathcal{EL}(t)$ affecting the system is described by means of a zero-mean Gaussian process ($\langle \mathcal{EL}(t)\rangle=0$). This is further characterized by auto-correlation function and is defined as $\mathcal{K}_{ORU}(t-t^{\prime},g)=\dfrac{g e^{-g|t-t^{\prime}|}}{2}$. To involve classical noise in the dynamical map of the four qubit state, we include the $\beta$-function, which reads as \cite{Yang-J, Rossi}:
\begin{equation}
\beta_{ORU}(t)=\int_0^t \int_0^t K(s-s)^{\prime}ds ds^\prime. \label{Beta function}
\end{equation}
Now, by inserting the auto-correlation function into Eq.\eqref{Beta function}, we get the final $\beta$-function for the ORU noise as \cite{Kenfack, Rossi}:
\begin{equation}
\beta_{{ORU}}(t)=\frac{1}{g}[g t+e^{-g t}-1],\label{Beta function of OU}
\end{equation}
where, $g$ is the inverse of the auto-correlation time and controls the memory characteristic of the classical environment. The characteristic function of the noise is given by $\langle \exp[\pm \iota n \varphi(t)]\rangle$=$ \langle \exp[-\dfrac{1}{2}n^2\beta_{ORU}(t)]\rangle$. In case of the CSE inetraction under the influence of ORU noise, we get the final density matrix by \cite{54, Kenfack, Rossi}:
\begin{equation}
\rho_{CSE}(t)= \left\langle U_{ijkl}(t)\rho_o U_{ijkl}(t)^{\dagger} \right\rangle_{\varphi_a}.\label{final density matrix of CSE}
\end{equation}
We generalize $\varphi_a=\varphi_b=\varphi_c=\varphi_d$. For the $BSE$ interaction, we obtain the final density matrix by:
\begin{equation}
\rho_{BSE}(t)= \left\langle \left\langle U_{ijkl}(t)\rho_o  U_{ijkl}(t)^{\dagger} \right\rangle_{\varphi_a} \right\rangle_{\varphi_d},\label{final density matrix of BSE}
\end{equation}
here, we assume $\varphi_b=\varphi_a$ and $\varphi_c=\varphi_d$. For the $TSE$ interaction situation, the corresponding final matrix takes the form as:
\begin{equation}
\rho_{TSE}(t)= \left\langle \left\langle \left\langle U_{ijkl}(t)\rho_o  U_{ijkl}(t)^{\dagger} \right\rangle_{\varphi_a} \right\rangle_{\varphi_b}\right\rangle_{\varphi_c},\label{final density matrix of TSE}
\end{equation}
Here, we assume $\phi_d=\phi_c$. In the final $ISE$ interaction, one can get the final density ensemble as:
\begin{equation}
\rho_{ISE}(t)= \left\langle \left\langle \left\langle \left\langle U_{ijkl}(t)\rho_o  U_{ijkl}(t)^{\dagger} \right\rangle_{\varphi_a} \right\rangle_{\varphi_b} \right\rangle_{\varphi_c} \right\rangle_{\varphi_d}.\label{final density matrix for ISE}
\end{equation}
here, $\langle ... \rangle$ shows taking average over all possible values of noise phase.
\section{Results and discussions}\label{Results and discussions}
In this section, we discuss the major results obtained from the analytical expressions evaluated for the dynamics of four qubit GHZ state within ORU noise-associated classical environments. Here, we analyze the role of ORU noise in four different system-environment coupling interactions whose final density ensembles are defined in Eqs.\eqref{final density matrix of CSE}, \eqref{final density matrix of BSE}, \eqref{final density matrix of TSE} and \eqref{final density matrix for ISE}. We mainly focus to determine the entanglement, coherence, and quantum information preservation by the system.
\subsection{Time evolution of four qubit GHZ state in CSE interaction}
We analyze dynamics of the four qubits when exposed to a single classical fluctuating environment described by ORU noise. By operating the entanglement witness, purity and entropy measures from Eqs.\eqref{ewo}, \eqref{purity} and \eqref{decoherence} on the final density matrix for the CSE interaction in Eq.\eqref{final density matrix of CSE}, the corresponding analytical expressions are as followed:
\begin{align}
EW_{CSE}(t)=&\frac{1}{32} \left(3+\mu_1+12 \mu_2\right),\\
P_{CSE}(t)=&\frac{1}{32} \left(19+\mu_1^{2}+12 \mu_2^2\right),\\
\mathcal{H}_{CSE}(t)=&-\mathcal{O}_1-\frac{1}{16} (7+\mathcal{O}_2) \log [\frac{1}{16} (7+\mathcal{O}_2) ]-\frac{1}{16} (7+\mathcal{O}_3) \log[\frac{1}{16} (7+\mathcal{O}_3)].
\end{align}
\begin{figure}[ht]
\centering
\includegraphics[scale=0.1]{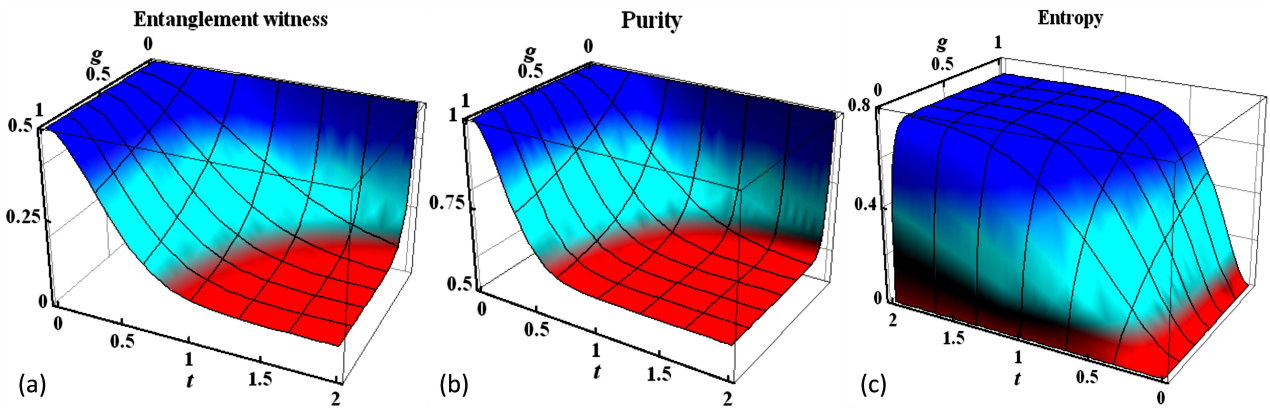}
\caption{Time-evolution of the entanglement witness (a), purity (b) and entropy (c) for four qubit GHZ state within common system-environment interaction described by Ornstein-Uhlenbeck noise when $g=1$ with time parameter $t=2$.}\label{dynamics of GHZ state under CSE}
\end{figure}
Fig.\ref{dynamics of GHZ state under CSE} shows the dynamics of $EW(t)$, $P(t)$ and $\mathcal{H}(t)$ evaluating preservation of entanglement, coherence, and quantum information for the system of maximally entangled four qubits in CSE interaction under $ORU$ noise. Because of the dephasing effects of the $ORU$ noise, we noticed that both $EW(t)$ and $P(t)$ decrease in time. The $\mathcal{H}(t)$, on the other hand, is realized to be the increasing quantum information decay function. Both statements demonstrate that the superposition of the noise phase over the system phase induces entanglement, coherence, and information decay, which were initially encoded in the four qubits. The noise strongly suppresses the system's characteristic fluctuating behavioural dynamics, leaving it with a pure monotonic structure. Thus, no entanglement sudden death (ESD) or birth (ESB) revivals have been observed. It means the CSE reduces the functionality of the system by not allowing the backflow of the information to the system. Hence, any kind of loss in the current case will be permanent with no recovery. The correlations between the qubits appear to be fragile, however, are not lost completely. As for $EW(t)$, the slope shows a smaller decline, determining smaller decay while not reaching the separability bound. Similarly, the slope of $P(t)$ does not reach the lower bound, indicating that the system remains coherent. Finally, the $\mathcal{H}(t)$ show that the four qubits undergo quantum information loss, however, in good correspondence with the $EW(t)$ and $P(t)$, one can deduce that the information is also not lost completely. Comparing the maxima and minima throughout comparable times reveals the correspondence between the measures. As the value of $g$ increases, the correlations between the qubits decrease more. Thus, decreasing $g$ improves the memory properties and correlations between the subsystems of the composite state under $ORU$ noise. In \cite{Kenfack}, using quantum negativity and quantum discord for the dynamics of a three-qubit state produced the same detrimental result, however, with different decay amounts. Under static and random telegraph noises, the dynamics of quantum negativity and discord of bipartite and tripartite states remained more stable and less reduced, demonstrating large-scale preservation with visible ESD and ESB phenomena \cite{Benedetti, Arthur, Kenfack-mixed}. In particular, we found that the current four-qubit entangled state is more optimal for quantum correlations, coherence, and information preservation than the bipartite one described in \cite{Rossi}.
\subsection{Time evolution of four qubit GHZ state in BSE interaction}
This section discusses the preservation of entanglement, coherence, and quantum information by deploying entanglement witness, purity, and entropy measure given in Eqs.\eqref{ewo}, \eqref{purity} and \eqref{decoherence} when two pairs of qubits are coupled independently to two classical environments with ORU noise. The final density ensemble state of the current BSE interaction is given in Eq.\eqref{final density matrix of BSE}.The following are the analytical expressions for the related measures:
\begin{align}
EW_{BSE}(t)=&\frac{1}{16} \zeta_1 \left(1+\zeta_2+\zeta_3+\zeta_4-\zeta_5 \right),\\
P_{BSE}(t)=&\frac{1}{16} \left(5+\zeta_6+\zeta_7+8 \zeta_8+\zeta_9\right),\\
\begin{split}
\mathcal{H}_{BSE}(t)=&-\frac{1}{8} \mathcal{P}_1 \log[\frac{1}{8} \mathcal{P}_2]-\frac{1}{8} \mathcal{P}_3\log[\frac{1}{8} \mathcal{P}_3]-\frac{1}{8} \mathcal{P}_4 \log[\frac{1}{8}\mathcal{P}_4]\\&-\frac{1}{32} \mathcal{P}_5\log[\frac{1}{32} \mathcal{P}_5]-\frac{1}{32}\mathcal{P}_6\log[\frac{1}{32}\mathcal{P}_6].
\end{split}
\end{align}
\begin{figure}[ht]
\centering
\includegraphics[scale=0.1]{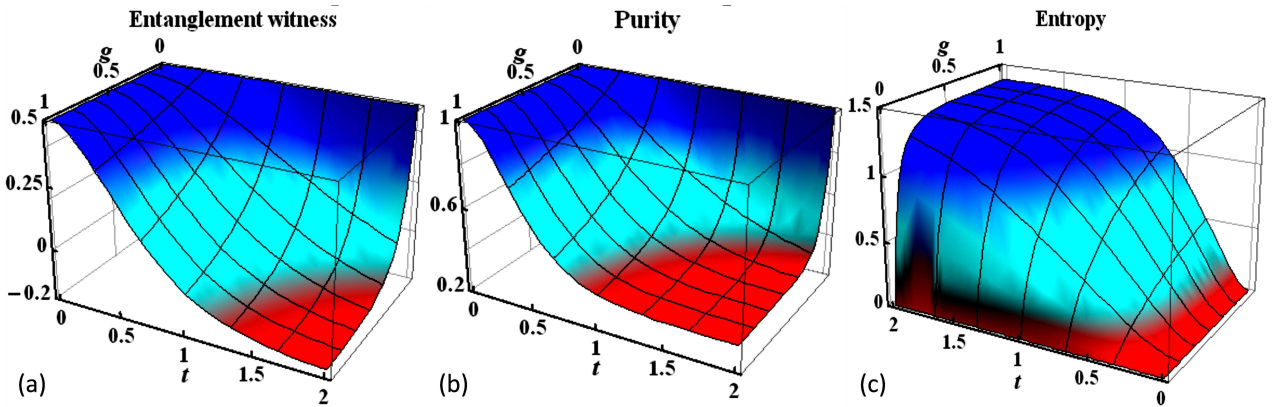}
\caption{Time-evolution of the entanglement witness (a), purity (b) and entropy (c) for four qubit GHZ state within bipartite system-environment interaction described by Ornstein-Uhlenbeck noise when $g=1$ with time parameter $t=2$.}\label{dynamics of GHZ state under BSE}
\end{figure}
In Fig.\ref{dynamics of GHZ state under BSE}, we explore the dynamics of $EW(t)$, $P(t)$ and $\mathcal{H}(t)$ in BSE interaction under ORU noise in four qubit GHZ state. In close alliance with the CSE, the BSE interaction also left the $EW(t)$ and $P(t)$ decreasing functions in time. The $\mathcal{H}(t)$ is found an increasing function of information decay. The qualitative dynamics shown by the measures are in good connection, displaying maximum and minimum values on the relative comparable time scales. However, the disentanglement rate of the four qubits quantitatively follows the coherence decay. Since coherence is a condition when it is lost, entanglement also disappears. Pure exponential decay drives the overall dynamical map, with no revivals observed. As a result, classical fluctuating environments are incompatible with the noise of a Gaussian probability distribution function, and their potential oscillating efficiency is lost. This was discovered to be the primary cause of the short-lived quantum correlations, coherence, and quantum information in the existence of ORU noise. This result is completely contradictory to the results obtained for noises with non-Gaussian probability distribution functions, such as static, random telegraph, and coloured noises \cite{Benedetti, Arthur, Kenfack-mixed, Benedetti-1f}. Not in agreement with CSE interaction, here, the system becomes completely decoherent and, as a result, disentangled according to the $P(t)$ and $EW(t)$. This does not guarantee complete quantum information decay, however, in comparison to the $CSE$, the current decay is larger according to $\mathcal{H}(t)$. At the upper limit of $g$, the decay becomes stronger, implying a non-supporting character for the entanglement, coherence, and quantum information preservation. There were no ESD or ESB phenomena found during the decay, which was characterized by pure monotonous functions in time. Since this phenomenon does not exist, there is no backflow of lost information. A similar decay aspect was observed by employing environments in pure and mixed noise cases for bipartite and tripartite states \cite{Kenfack, Rossi,ff}. Under non-Gaussian noises, however, the complete dynamical map, as well as the preservation capabilities of quantum systems, becomes increasingly different \cite{Benedetti, Arthur, Kenfack-mixed}. The corresponding maximum and minimum of $EW(t)$ and $P(t)$ are in close agreement with that of the $\mathcal{H}(t)$, meaning that the results are consistent and correct.
\subsection{Time evolution of four qubit GHZ state in TSE interaction}
Here, we address the dynamics of entanglement witness, purity and entropy for the four qubit system when coupled to three independent environmental sources described by ORU noise. In this scheme, we couple two qubits to one, while the other two qubits are coupled separately to two other environments. For this configuration, the corresponding final density matrix is given by Eq.\eqref{final density matrix of TSE}. Now, by using the Eqs.\eqref{ewo}, \eqref{purity}, \eqref{decoherence}, we give the numerical results obtained for the entanglement, coherence and quantum information are as followed:
\begin{align}
EW_{TSE}(t)=&\frac{1}{16} \left(-5+\omega_1+\omega_2+11 \omega_3\right),\\
P_{TSE}(t)=&\frac{1}{16} \omega_4 \left(1+\omega_5+11 \omega_6+3 \omega_7\right),\\
\begin{split}
\mathcal{H}_{TSE}(t)=&-\frac{1}{4} \mathcal{Q}_1\log[\frac{1}{4} \mathcal{Q}_1]-\frac{1}{8} \mathcal{Q}_2\log[\frac{1}{8} \mathcal{Q}_2]-\frac{1}{8} \mathcal{Q}_3\log[\frac{1}{8} \mathcal{Q}_3]\\
&-\frac{1}{8} \mathcal{Q}_4\log[\frac{1}{8} \mathcal{Q}_4]-\frac{1}{16} \mathcal{Q}_5\log[\frac{1}{16} \mathcal{Q}_5]-\frac{1}{16} \mathcal{Q}_6\log[\frac{1}{16} \mathcal{Q}_6].
\end{split}
\end{align}
\begin{figure}[ht]
\centering
\includegraphics[scale=0.1]{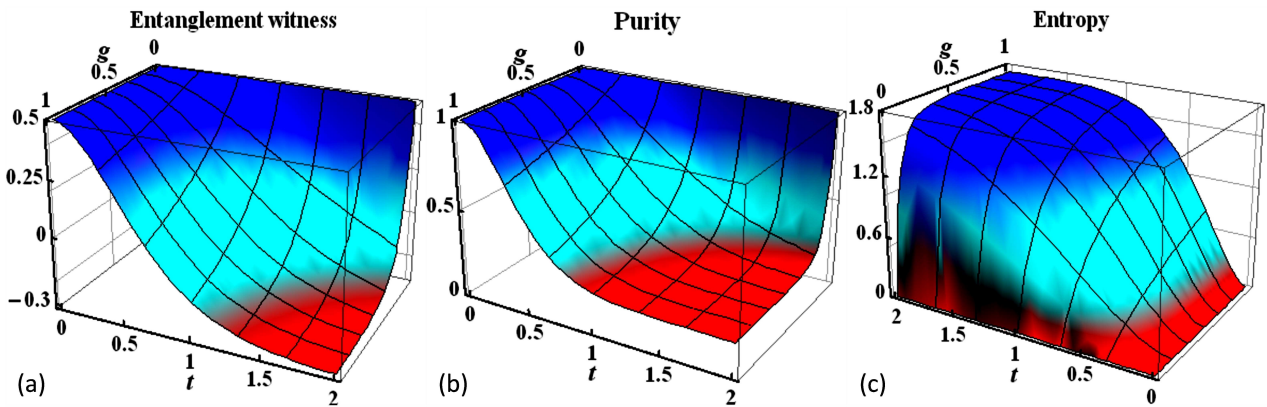}
\caption{Time-evolution of the entanglement witness (a), purity (b) and entropy (c) for four qubit GHZ state within tripartite system-environment interaction described by Ornstein-Uhlenbeck noise when $g=1$ with time parameter $t=2$.}\label{dynamics of GHZ state under TSE}
\end{figure}
Fig.\ref{dynamics of GHZ state under TSE} the preservation of entanglement, coherence, and quantum information under ORU noise in TSE interaction is discussed. The $EW(t)$, $P(t)$ are found decreasing functions of entanglement and coherence in time, whereas, the $\mathcal{H}(t)$ has shown an increase in the information decay. From the results, one can immediately uncover that the decay encountered here is greater than those observed in CSE and BSE interaction shown in Figs. \ref{dynamics of GHZ state under CSE} and \ref{dynamics of GHZ state under BSE}. The coupling of the system with many environments is the primary cause of this decay. This means that when a system is coupled with many environments, quantum correlations, coherence, and quantum information all become weaker, and in most situations, absolute separability is reached. According to the $EW(t)$ and $P(t)$, the entanglement and coherence undergo greater decay than that observed in the previous cases. In agreement, the $\mathcal{H}(t)$ reaches a higher saturation level, showing a greater loss of quantum information. Qualitatively, the dynamical map is shown by the measures for the entanglement, coherence, and quantum information resemble. This affirms that there is an intrinsic relation between the three phenomena and are interconnected. However, the coherence decay seems faster than the entanglement loss for the system. It means that a rise in the decoherence between the system and TSE triggered the disentanglement of the system. The decay observed is completely monotonic with no ESD and ESB phenomenon. Thus, lacking the exchangeability of the information between the system and environment resulting in permanent information loss. Except for the qualitative nature of the decay, we notice that the current quantitative results differ from those published in \cite{Kenfack} with the same noise for the tripartite entangled state. In the case of quantum correlations, coherence, and information preservation, this means that various quantum systems may be affected differently by the same noise. Besides, all the three measures have provided the same results, showing the strict agreement between them. The investigation for entangled states revealed completely different results when subjected to classical environments with non-Gaussian noises \cite{Benedetti, Arthur, Kenfack-mixed, Benedetti-1f}. When the current four-qubit system is exposed to three independent environments, the degree of deterioration is larger than when three qubits are subjected to three independent environments \cite{Kenfack}. Thus, the system's fragility is influenced not only by the noise but also by the number of environments and quantum system involved.
\subsection{Time evolution of four qubit GHZ state in ISE interaction}
The dynamics of entanglement, coherence with quantum information preservation for the system of four qubits is given in detail. In the current configuration, each one of the four qubits is coupled to four separate classical environments. The final density matrix state of the ISE interaction is given in Eq.\eqref{final density matrix for ISE}. For the numerical simulations obtained from the entanglement witness, purity and entropy by using the Eqs. \eqref{ewo}, \eqref{purity} and \eqref{decoherence} are followed as:
\begin{align}
EW_{ISE}(t)=&\frac{1}{8} \left(-3+\kappa_1+6 \kappa_2\right),\\
P_{ISE}(t)=&\frac{1}{4} \kappa_3 (3+\cosh[\kappa_4 ]),\\
\begin{split}
\mathcal{H}_{ISE}(t)=&-\frac{1}{2} \mathcal{R}_1 \log\left[\frac{1}{8} \mathcal{R}_1\right]-\frac{3}{8} \mathcal{R}_2 (\mathcal{R}_3)^2 \log\left[\frac{1}{8} \mathcal{R}_2 (\mathcal{R}_3)^2\right]\\&-\frac{1}{8} \mathcal{R}_4 (1+\mathcal{R}_5) \log\left[\frac{1}{8} \mathcal{R}_4(1+\mathcal{R}_5)\right].
\end{split}
\end{align}
\begin{figure}[ht]
\centering
\includegraphics[scale=0.1]{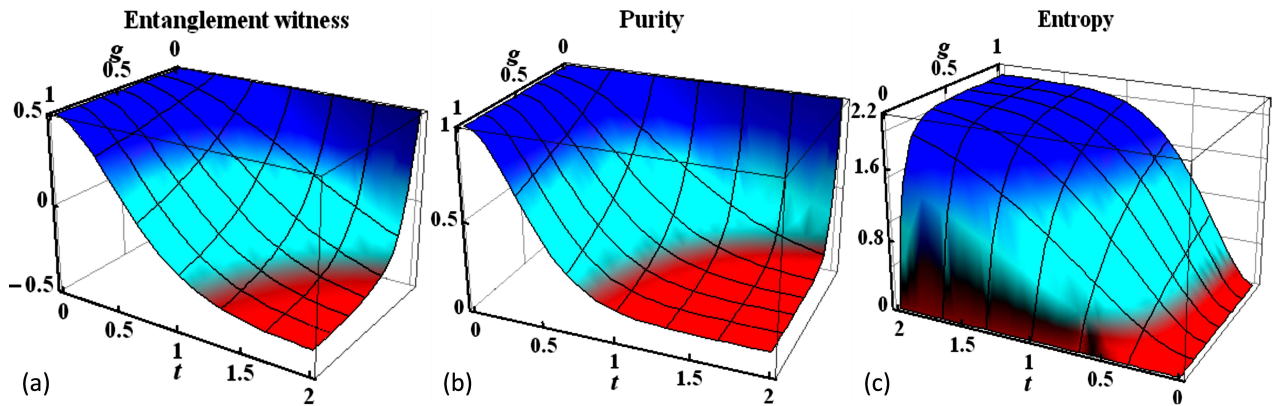}
\caption{Time-evolution of the entanglement witness (a), purity (b) and entropy (c) for four qubit GHZ state within tripartite system-environment interaction described by Ornstein-Uhlenbeck noise when $g=1$ with time parameter $t=2$.}\label{dynamics of GHZ state under ISE}
\end{figure}
Fig.\ref{dynamics of GHZ state under ISE} evaluates the time-dependent dynamics of the $EW(t)$, $P(t)$ and  $\mathcal{H}(t)$ measures within ISE interaction under ORU noise. In agreement with the previous results, both $EW(t)$ and $P(t)$ is found to reduce functions of entanglement and coherence, while $\mathcal{H}(t)$ remained an increasing function of quantum information decay. In comparison, the current detrimental effects of the four environments are much greater than in all previous cases. This is relatively the larger encountered decay for various types of quantum systems, environments, and situations under various Gaussian and non-Gaussian noises \cite{33,35,36,45,46, Kenfack, Rossi}. No ESD and ESB phenomenon has been observed and the decay that occurred was pure monotonic with no revivals. Because of this, entanglement, coherence, and quantum information remained short-lived. In previous studies, due to the entanglement oscillations, longer preserved quantum correlations employing the quantum negativity and quantum discord were detected \cite{Benedetti, Arthur, Tchoffo-2, Tsokeng}. This also confirms that there is no information backflow mechanism between the system and environment, resulting in a permanent loss. Moreover, for the GHZ state, the coherence seems preserved for smaller intervals than the entanglement. This is because of the loss of coherence between the system and ISE, which results in complete separability and huge quantum information loss. This is in contrast to the quantum correlations preservation capacities of the entangled state in independent environments under the non-Markovian regime of non-Gaussian noises as described in \cite{Benedetti, Arthur, Kenfack-mixed, Kenfack}. All three measures agree with one another and provide the same qualitative dynamical map, ensuring that the results are consistent. In agreement with previous results, for the increasing values of $g$, the entanglement, coherence, and information decay tend to increase. Thus, the smaller values of $g$ can be helpful to produce a longer preservation capacity. However, the detrimental effects in the current ISE interaction under ORU noise seem unavoidable and the state becomes separable in any case. The degradation level observed in four independent environments is far higher than that observed in three independent environments under the same noise \cite{Kenfack}.
\subsection{Detail dynamics against different values of noise parameter $g$}
The present section address the dynamics of the entanglement, coherence and quantum information initially maximally encoded in GHZ state exposed to classical fluctuating environments with ORU noise. Most of this section is dedicated to a thorough approach of time evolution for various values of $g$.
\subsubsection{Entanglement, coherence and information preservation in CSE interaction}
\begin{figure}[ht]
\centering
\includegraphics[scale=0.1]{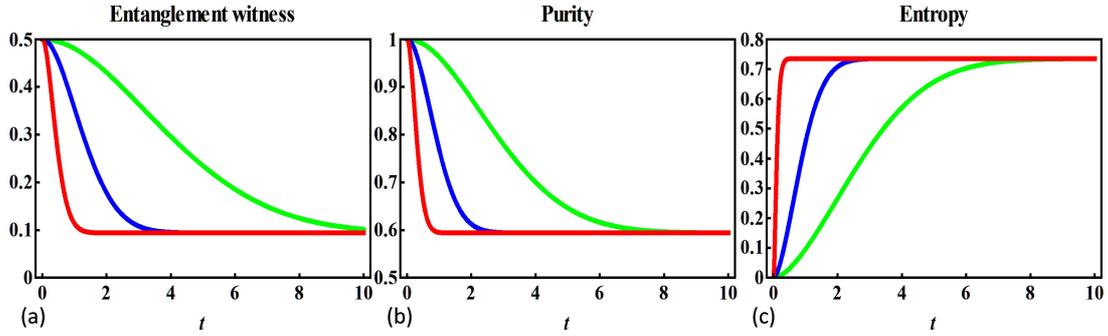}
\caption{Time-evolution of the entanglement witness (a), purity (b) and entropy (c) for four qubit GHZ state within common system-environment interaction described by Ornstein-Uhlenbeck noise when $g=10^{-2}$ (green), $g=10^{-1}$ (blue), $g=10$ (red) with time parameter $t=10$.}\label{detail dynamics of GHZ state under 	CSE}
\end{figure}
In Fig.\ref{detail dynamics of GHZ state under CSE}, we analyze the time evolution process for the GHZ state in CSE interaction for different values of $g$. The detrimental dephasing effects of the ORU noise caused the initially encoded entanglement, coherence, and quantum information to degrade. In CSE interaction, the state is not entirely separable; however, the degrading effects of noise cannot be avoided. We found $g$ to have a significant impact on the preservation capacity of the GHZ state in classical random fields. The slopes shift from green towards the red-end for the increasing values of $g$, indicating an inverse relation with the preservation time. Thus, by reducing the noise parameter $g$, one can strengthen entanglement, coherence, and quantum information preservation. According to the measures, long-term preservation is detected for $g=10^{-2}$ relatively. This detrimental behaviour of $g$ entirely contradicts the memory properties of the Hurst exponent of the fractional Gaussian noise \cite{Rossi}. We found the GHZ state more preservative for the entanglement, coherence, and quantum information according to the $EW(t)$, $P(t)$ and $\mathcal{H}(t)$. This is consistent with the previous results shown in Fig.\ref{dynamics of GHZ state under CSE}. Thus, utilizing the GHZ state in a single environment for quantum information processing applications remains one of the best choices where entanglement preservation is a required feature. No ESD and ESB revivals have been observed, and the decay is smooth monotonic. This comprehends the no backflow mechanism for information between the system and CSE and results in permanent decay rather than temporary loss, for example as shown in \cite{Benedetti, Arthur, Kenfack-mixed, Tchoffo-2}. The measures showed a one-to-one correspondence between the maxima and minima, assuring good qualitative connections between them and resulting in accuracy and validity. In the quantitative analysis, the decoherence rate occurs faster than the disentanglement rate of the system. In the current case, the overall decay detected is considerably different from those defined in \cite{Benedetti, Arthur, Kenfack-mixed, Tsokeng,44, Benedetti-1f, Benedetti}. Most importantly, we found the four qubit GHZ state more suitable for the quantum correlations, coherence, and information preservation than the bipartite Bells state and mixture states investigated under fractional Gaussian and ORU noise given in \cite{Rossi}.
\begin{table}[ht]
\centering
\caption{Shows saturation levels, saturation time and $\beta$-functions for corresponding values of $g$ for CSE interaction.}\label{table-CSE}
\begin{tabular}{|c|c|c|c|c|c|c|c|c|}
\hline
g 		 &$EW(t)$& $S.T$ & $P(t)$ & $S.T$  & $\mathcal{H}(t)$ & $S.T$  &$\beta$-function\\
\hline
$10^{-2}$& 0.09	& 10 	&0.6  & 9.0	&0.73 		&9.0 & 	$a_{i,j}(\frac{100}{e^{1/10}}-90$)\\
\hline
$10^{-1}$& 0.09	& 4.5	& 0.6 & 3.0	&0.73		&3.0 & $a_{i,j}(\frac{10}{e})$\\
\hline
$10$   	 & 0.09 & 1.5	&0.6  & 1.3	&0.73		&0.8 &	$a_{i,j}(9.9+\frac{1}{10e^{100}})$\\
\hline
\end{tabular}
\end{table}
In table \ref{table-CSE}, saturation levels are are indicated by $EW(t)$, $P(t)$ and $\mathcal{H}(t)$. The saturation time, where the first death of the entanglement, coherence and information occurs is shown by $S.T$ for each measure. For each value of $g$, we also indicated the value of $\beta_{ORU}$ and we noticed that for a fixed values of $t$, if $\beta_{ORU}$ increases, the corresponding $S.T$ decreases. In addition, $a_{i,j}$ for the CSE interaction equals $4$ and $8$.
\subsubsection{Entanglement, coherence and quantum information preservation for GHZ state in BSE interaction}
\begin{figure}[ht]
\centering
\includegraphics[scale=0.1]{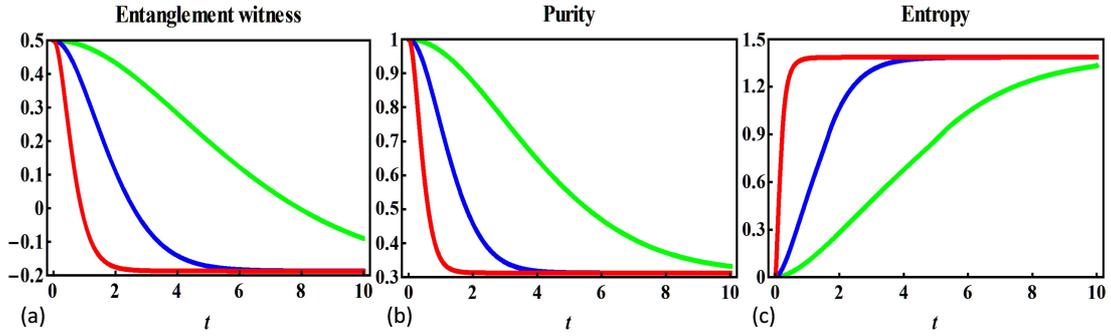}
\caption{Time-evolution of the entanglement witness (a), purity (b) and entropy (c) for four qubit GHZ state within bipartite system-environment interaction described by Ornstein-Uhlenbeck noise when $g=10^{-2}$ (green), $g=10^{-1}$ (blue), $g=10$ (red) with time parameter $t=10$.}\label{detail dynamics of GHZ state under BSE}
\end{figure}
The time-dependent dynamics of the entanglement, coherence, and quantum information is evaluated under ORU noise in BSE interaction in Fig.\ref{detail dynamics of GHZ state under BSE}. We find the entanglement, coherence, and quantum information more fragile than that at CSE interaction and undergo a larger decay. The slopes attain a saturation level below the lower bound of the $EW(t)$ and $P(t)$, thus, suggesting that the system becomes decoherent and disentangled. In agreement with the previous results, the disentanglement speed lags behind the coherence decay. This affirms the fact that it causes the entanglement lost by the system when the system and surroundings become out of phase. Moreover, for the increasing values of $g$, we found the slopes tend to shift from green towards the red end, showing larger decay. Besides, the random values of $g$ influenced the preservation of time. Thus, for the quantum mechanical protocols requiring greater information preservation, the relevant environment should be described by the smaller values of $g$. However, in BSE interaction, the noise quickly superimposes the system's phase. Hence, unlike in the case of CSE interaction, the ORU noisy effects in BSE interaction are inescapable, and the state in any condition will become separable. The decay observed is pure monotonic, with no evidence of entanglement oscillations. This suggests that the ORU noise causes the classical environments to lose their fluctuating behaviour. This is incompatible with the majority of quantum computing activities, resulting in hardware and application performance issues. The significance of the fluctuating behaviour is given in detail in \cite{Benedetti, Arthur, Kenfack-mixed, Tchoffo-2, Benedetti-1f, Benedetti-CN}, where due to ESD and ESB phenomena, quantum correlations were preserved for longer periods in time. Besides, all the measures have produced the same qualitative results, showing good connections between them. We observed that the current decay amounts in the BSE interaction are similar to those found in independent environments under ORU noise for the tripartite state \cite{Kenfack}. Although, it is notable that the nature of both the environmental set-ups differ.\\
\begin{table}[ht]
\centering
\caption{Shows saturation levels, saturation time and $\beta$-functions for corresponding values of $g$ for BSE interaction.}\label{table-BSE}
\begin{tabular}{|c|c|c|c|c|c|c|c|c|}
\hline
g 		 &$EW(t)$& $S.T$ & $P(t)$ & $S.T$  & $\mathcal{H}(t)$ & $S.T$  &$\beta$-function\\
\hline
$10^{-2}$&-0.2& $10^{\prime}$&0.32&$10^{\prime}$ &1.35 		&$10^{\prime}$ & 	$b_{i,j}(\frac{100}{e^{1/10}}-90)$\\
\hline
$10^{-1}$& -0.2	& 7.0	& 0.32 & 5.0	&1.35		&5.0 & $b_{i,j}(\frac{10}{e})$\\
\hline
$10$   	 & -0.2 & 3.0	&0.32  & 2.3	&1.35		&1.3 &	$b_{i,j}(9.9+\frac{1}{10e^{100}})$\\
\hline
\end{tabular}
\end{table}
Note that, $10^{\prime}$ means saturation time greater than $10$ with $b_{i,j}=2,4$.
\subsubsection{Entanglement, coherence and quantum information preservation in TSE interaction}
\begin{figure}[ht]
\centering
\includegraphics[scale=0.1]{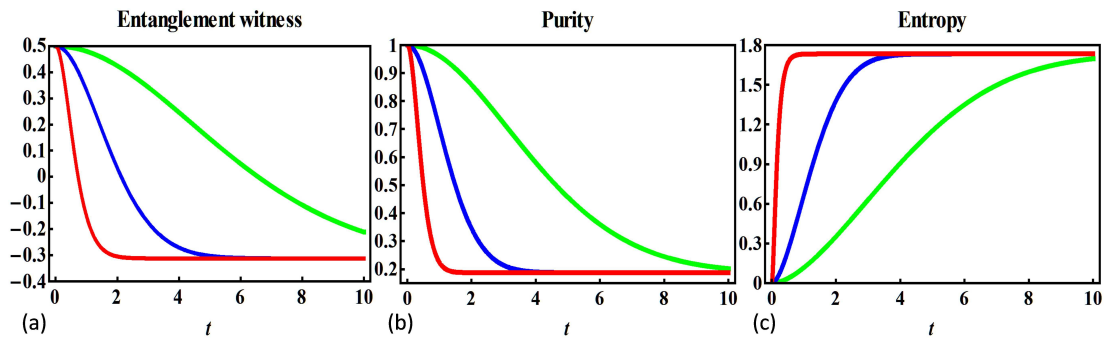}
\caption{Time-evolution of the entanglement witness (a), purity (b) and entropy (c) for four qubit GHZ state within tripartite system-environment interaction described by Ornstein-Uhlenbeck noise when $g=10^{-2}$ (green), $g=10^{-1}$ (blue), $g=10$ (red) with time parameter $t=10$.}\label{detail dynamics of GHZ state under TSE}
\end{figure}
Fig.\ref{detail dynamics of GHZ state under TSE} address the time-evolution of the $EW(t)$, $P(t)$ and $\mathcal{H}(t)$ when two qubits are coupled to one environment, while the other two are coupled to two independent environments, with ORU noise. We found the ORU noise more detrimental where the final saturation levels indicate greater entanglement, coherence, and information decay than observed in Figs.\ref{detail dynamics of GHZ state under 	CSE} and \ref{detail dynamics of GHZ state under BSE} for CSE and BSE interactions. This decay, in comparison, is greater than detected for the tripartite state in both mixed and independent kinds of interactions, for example as studied in \cite{Kenfack}. Except for the CSE interaction in the current analysis, we deduce that the four qubit states, in addition to having more information storage capacity, are more vulnerable to external noisy effects than the tripartite GHZ state \cite{Kenfack}. Furthermore, it is clear that the parameter $g$ has a significant impact on preservation, and that for smaller values, the decay is reduced directly. Regardless, the entanglement and coherence dissipate entirely, and the state becomes separable, eventually. This contrasts the findings for bipartite and tripartite states under various types of system-environment coupling and noises, where entanglement lasts longer \cite{Kenfack-mixed, Benedetti-1f, Benedetti,ff, Tchoffo-2, dd}. Besides, one can note that the decay amount varies in each system-environment coupling, which affirms that the decay is not only described by the noise but also by the type of involved system-environment coupling. In addition, the $EW(t)$, $P(t)$ and $\mathcal{H}(t)$ all are monotonic functions in time with no ESD and ESB phenomenon. This suggest that the nature of decay is only governed by the type of noise involved and not by the type of system-environments interaction. As different types of decay have been observed in the same classical environments, however, with distinct noises given in \cite{dd, Tchoffo-2, Benedetti, ff, gg, Benedetti-1f}. Note that the decay amount is mostly controlled by the type of system-environment coupling, as for different coupling schemes, one can observe different decay amounts and preservation intervals. In agreement with the previous findings, coherence is decaying faster than the entanglement assuring different natures of the phenomena. All the measures provided the same qualitative results, showing a close alliance between them by showing correspondence between their maxima and minima.\\
\begin{table}[ht]
\centering
\caption{Shows saturation levels, saturation time and $\beta$-functions for corresponding values of $g$ for TSE interaction.}\label{table-TSE}\begin{tabular}{|c|c|c|c|c|c|c|c|c|}
\hline
g 		 &$EW(t)$& $S.T$ & $P(t)$ & $S.T$  & $\mathcal{H}(t)$ & $S.T$  &$\beta$-function\\
\hline
$10^{-2}$&-0.3& $10^{\prime}$&0.18&$10^{\prime}$ &1.73 		&$10^{\prime}$ & 	$c_{i,j}(\frac{100}{e^{1/10}}-90)$\\
\hline
$10^{-1}$& -0.3	& 7.0	& 0.18 & 4.5	&1.73		&4.5 & $c_{i,j}(\frac{10}{e})$\\
\hline
$10$   	 & -0.3 & 3.0	&0.18  & 1.8	&1.73		&1.8 &	$c_{i,j}(9.9+\frac{1}{10e^{100}})$\\
\hline
\end{tabular}
\end{table}
Here, $b_{i,j}=2,4$.
\subsubsection{Entanglement, coherence and information preservation in ISE interaction}
\begin{figure}[ht]
\centering
\includegraphics[scale=0.1]{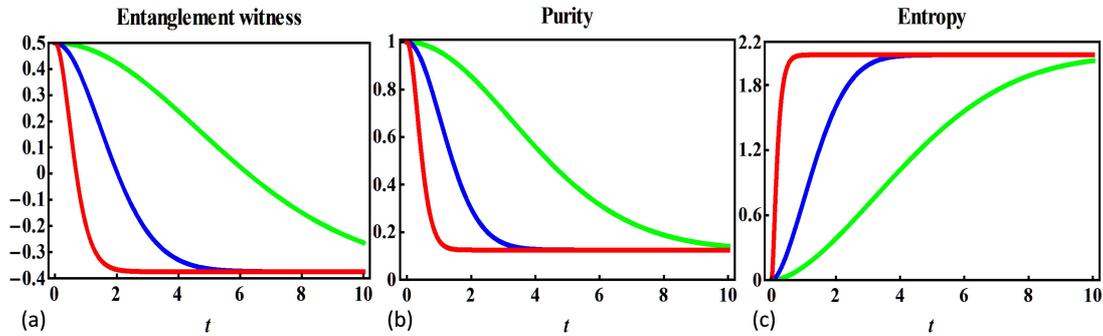}
\caption{Time-evolution of the entanglement witness (a), purity (b) and entropy (c) for four qubit GHZ state within independent system-environment interaction described by Ornstein-Uhlenbeck noise when $g=10^{-2}$ (green), $g=10^{-1}$ (blue), $g=10$ (red) with time parameter $t=10$.}\label{detail dynamics of GHZ state under ISE}
\end{figure}
In Fig.\ref{detail dynamics of GHZ state under ISE}, dynamics of the $EW(t)$, $P(t)$ and $\mathcal{H}(t)$ for the four qubit GHZ state when coupled to ISE is analyzed. We found the ISE model the most detrimental, out of all the current configurations. As the saturation levels for all three measures indicate greater decay in comparison. Besides, the dephasing effects in the ISE interaction in the four-qubit state are also found much greater than that for the three-qubit state of the same kind and noise \cite{Kenfack}. This indicates that the quantum correlations, coherence, and quantum information decay have an intrinsic relationship with the number of independent environments and, for the increasing number of environments, the decay rate directly increases. Hence, besides the advantage of many channel capacities in the current scheme, the decay amount is significantly large and should be avoided in the quantum mechanical processes, requiring greater efficiency. For the increasing values of $g$, the slopes shift from the lower decay end towards the greater decay end.
\begin{table}[ht]
\centering
\caption{Shows saturation levels, saturation time and $\beta$-functions for corresponding values of $g$ for ISE interaction.}\label{table-ISE}
\begin{tabular}{|c|c|c|c|c|c|c|c|c|}
\hline
g 		 &$EW(t)$& $S.T$ & $P(t)$ & $S.T$  & $\mathcal{H}(t)$ & $S.T$  &$\beta$-function\\
\hline
$10^{-2}$&-0.37& $10^{\prime}$&0.13&$10^{\prime}$ &2.1 		&$10^{\prime}$ & 	$2(\frac{100}{e^{1/10}}-90)$\\
\hline
$10^{-1}$& -0.37	& 6.7	& 0.13 & 4.5	&2.1		&4.5 & $2(\frac{10}{e})$\\
\hline
$10$   	 & -0.37 & 2.5	&0.13  & 1.8	&2.1		&1.3 &	$2(9.9+\frac{1}{10e^{100}})$\\
\hline
\end{tabular}
\end{table}Thus, $g$ has anti memory preservation effects and influences the entanglement, coherence, and quantum information degradation. This deterioration capability of $g$ does not meet with the properties of the memory parameter of the fractional Gaussian noise \cite{49}. With no entanglement revivals, the overall decay is characterized by decreasing monotonous functions in time. This suggests that the classical environments with ORU noise do not support ESD and ESB phenomena and the resulting decay will be permanent with no back recovery of the information. The results for the independent classical environments with non-Markovian nature, such as random telegraph, static, and coloured noises, are incongruent, with the state remaining entangled bearing entanglement revival properties, as can be observed in\cite{Benedetti, Arthur, Kenfack-mixed, Benedetti-1f, Benedetti-CN, gg, dd, ff}. As a result, we infer that the qualitative decay is largely determined by the type of noise present, whereas the amount of decay is primarily dictated by the type of system-environment interaction involved. The present dephasing behaviour of classical environments is distinct from that of other types of environments aided by baths and Jaynes-Cummings model \cite{32,33,35,36,44,45,46,47}.
\subsubsection{Comparative time evolution}
\begin{figure}[ht]
\centering
\includegraphics[scale=0.1]{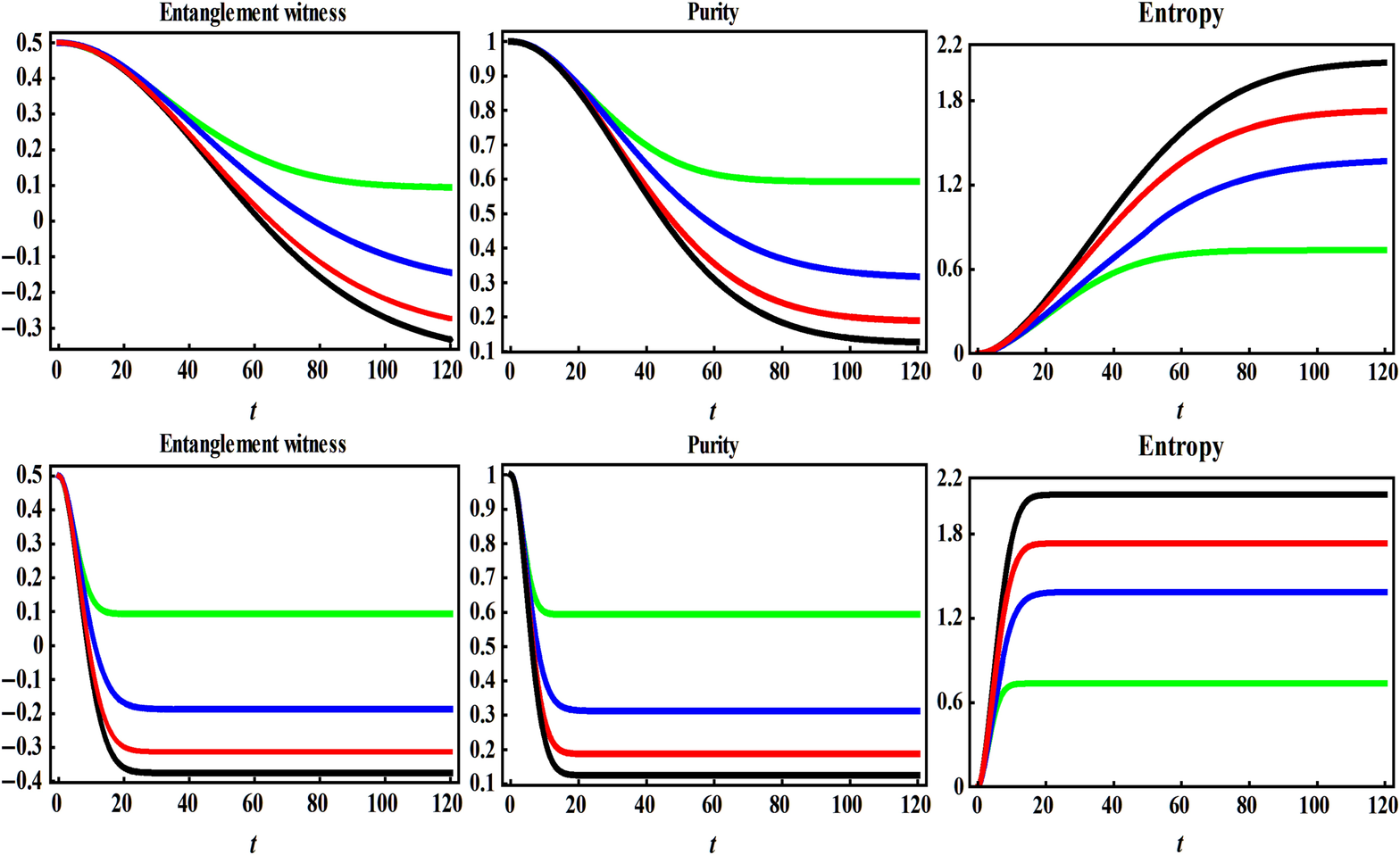}
\caption{Upper panel: Time-evolution of the entanglement witness (a), purity (b) and entropy (c) for four qubit GHZ state within common (green), bipartite (blue), tripartite (red) and independent system-environment (black) interaction described by Ornstein-Uhlenbeck noise when $g=10^{-4}$ with time parameter $t=120$. Bottom panel: same as upper panel but when $g=5 \times 10^{-3}$.}\label{comparative dynamics}
\end{figure}
The present section addresses the dynamics of entanglement, coherence, and quantum information initially maximally encoded in GHZ state exposed to classical fluctuating environments with ORU noise. Most of this section is dedicated to a thorough approach of the time evolution for different values of $g$. Besides, we intend to engineer long-lived preservation of the quantum phenomenon by using a wide range of the corresponding noise parameter.\\
Fig.\ref{comparative dynamics} shows a comparative time-dependent dynamical map of the entanglement, coherence, and quantum information, initially encoded in four qubit GHZ state when coupled to different classical environmental configurations under ORU noise. It's worth noting that the decay levels are at different elevations for the same values of $g$, implying that the dephasing behaviour of the noise is random across different classical environments. Besides, the system undergoes the least decay in CSE interaction. It is the only environment, according to the $EW(t)$ and $P(t)$, where the system remained entangled and coherent. In contrast, in all other cases, the system becomes fully decoherent and separable. However, the long-term entanglement, coherence, and quantum information cannot be ignored in all the configuration models for low values of $g$. This turns out to be more resourceful than in all other cases where the preservation time is shorter relatively \cite{Kenfack, Rossi, Benedetti, Arthur, Kenfack-mixed, 44, 45}. This suggests the versatility of the Gaussian distribution function of the ORU noise, which can enable the quantum practitioners to reduce the relative detrimental effects to optimal values for realistic quantum computing. This does not agree with the characteristic distribution function of various noises, which has a limited range of noise parameter values and is difficult to avoid. Static, fractional Gaussian and coloured noises are examples of this type of noise \cite{Arthur, Rossi, ff, Benedetti-1f}. Apart from the massive decay across various environments, the quantum information encoded in the system cannot be considered completely lost. For the increasing number of environments, the slopes are observed to shift from green towards the black end. This implies that the number of environments has a direct effect on entanglement, coherence, and quantum information dissipation. The current findings also suggest that no further decay occurs once the final saturation levels are reached. This is in contrast to the random telegraph, static, and coloured noises, which have all shown random behaviour, especially in non-Markovian regimes \cite{Benedetti, Arthur, Kenfack-mixed, Tchoffo-2, Benedetti-1f, ff, gg, dd}. Following that, the actual fluctuating character of the environments is lost because of the superposition of the noise phase over the joint phase of the system and environments. By not preserving the quantum phenomenon through their fluctuations, the effective performance of classical environments is thus reduced. This characteristic of classical environments has been observed to support the preservation capabilities described in \cite{Benedetti-1f, Benedetti, gg, dd, Tchoffo-2, Benedetti, Arthur}. By blocking the back-flow of information between the system and the environment, the practical implementation of environments with ORU noise will not allow the free states to convert into resource states. As a result, any decay found here will be permanent rather than temporary, as seen in \cite{Benedetti-1f, Benedetti-CN, Tchoffo-2, gg, ff, dd}.
\begin{table}[ht]
\centering
\caption{Shows saturation levels, saturation time and $\beta$-functions for corresponding values of $g$ for CSE,  BSE, TSE and ISE interactions.}\label{table-comparative}
\begin{tabular}{|c|c|c|c|c|c|c|c|c|c|}
\hline
Environment &  $g$ 	& $EW(t)$ 	& $S.T$ & $P(t)$ & $S.T$  & $\mathcal{H}(t)$ & $S.T$  & $\beta$-function &Remarks \\
\hline
CSE      &$ 10^{-4}$	& 0.09	& $110$ &0.6  & 90	&0.73 		&90 & $a_{i,j}\left(-9880+\frac{10000}{e^{3/250}}\right)$ &Least destructive\\ 
			& $5\times10^{-3}$	& 0.09	& 17	& 0.6	& 16	&0.73		&16	&$a_{i,j}\left(-80+\frac{200}{e^{3/5}}\right)$	&\\
\hline\hline

BSE      &$ 10^{-4}$	& -0.2	& $120^{\prime}$ &0.32  & 120	&1.35 		&120 & $b_{i,j}\left(-9880+\frac{10000}{e^{3/250}}\right)$ &\\ 
		& $5\times10^{-3}$	& -0.2	& 30	& 0.32	& 20	&1.35		&20	&$b_{i,j}\left(-80+\frac{200}{e^{3/5}}\right)$	&\\
\hline\hline
TSE      &$ 10^{-4}$	& -0.3	& $120$ &0.18 & 117	&1.73 		&117 & $c_{i,j}\left(-9880+\frac{10000}{e^{3/250}}\right)$ &\\ 
	& $5\times10^{-3}$	& -0.3	& 25	& 0.18	& 20	&1.73		&18	&$c_{i,j}\left(-80+\frac{200}{e^{3/5}}\right)$	&\\

\hline\hline
ISE      &$ 10^{-4}$	& -0.37	& $110$ &0.13  & 120	&2.1 		&120 & $d_{i,j}\left(-9880+\frac{10000}{e^{3/250}}\right)$ &Most destructive\\ 
		& $5\times10^{-3}$	& -0.37	& 17	& 0.13	& 23	&2.1		&20	&$d_{i,j}\left(-80+\frac{200}{e^{3/5}}\right)$	&\\
\hline\hline
\end{tabular}
\end{table}
\section{Conclusion}\label{Conclusion}
We have analyzed the dynamics of entanglement, coherence, and quantum information initially encoded in a maximally entangled four-qubit GHZ state when subjected to different classical fluctuating environments, namely: common, bipartite, tripartite, and independent environments described by Ornstein-Uhlenbeck (ORU) noise. In the first case, we connect the system with one classical environment, while in the second one, it is coupled to two. The four qubits are coupled to three environments in the tripartite case, and each qubit is coupled to four environments independently in the final case. Entanglement, coherence, and quantum information preservation have all been examined using entanglement witness, purity, and Shannon entropy.\\
Our findings suggest that the decay is highly controlled by the type of system-environment interaction and directly increases with the number of environments. We found the common environments to be the least dephasing and the only case, where the state remained partially entangled and coherent. In contrast, the state becomes completely separable and decoherent within all other configurations, which contradicts the results obtained under random telegraph and coloured noises \cite{Tchoffo-2, Benedetti-1f, Benedetti, ff}. The decay encountered in the case of the tripartite GHZ state differs considerably from that occurred for the four-qubit state of the same kind \cite{Kenfack}. In contrast to the easily separable bipartite Bell's and mixture states under ORU noise, the four qubit GHZ state remained entangled and coherent assuring its effectiveness relatively \cite{Rossi}.\\
The superposition of the noise phase over the system phase has completely suppressed the fluctuating behaviour of the classical environments, which resulted in pure monotonic decay, with no entanglement oscillations. This would also block transitions of the free states into resource states during the functional implementations of the quantum protocols \cite{ESD and ESB}. This indicates the Markovian character of the current environments for the quantum correlation, coherence, and information dynamics.\\
Moreover, along with the required selection of the system-environment coupling, the proper control of the noise parameters is also found important for the preservation effects. We found the entanglement, coherence, and quantum information preserved for a longer duration, particularly when $g=10^{-4}$. This is long enough compared to those given in \cite{Benedetti, Arthur, Kenfack-mixed, 44, 45, Benedetti-1f, gg, dd, Tchoffo-2, Benedetti, ff, Kenfack, Rossi} for different quantum systems, noises and environments. Unlike other noises, such as fractional Gaussian noise, the ORU noisy spectrum has been found flexible, allowing the environments to preserve entanglement, coherence, and information for longer periods with improved memory features nearly in all the configurations \cite{54, Rossi,57}. The computational statistics shows that the the $\beta$-function in each system-environment interaction is different. From tables \ref{table-CSE}, \ref{table-BSE}, \ref{table-TSE}, \ref{table-ISE} and \ref{table-comparative}, we found that at fixed values of $t$, the $\beta$-functions directly increases with $g$ and produces larger decay proportionally.\\
Our results also confirm the different natures of entanglement and coherence. The information encoded in the GHZ state is found to be relatively more dependent upon the coherence than the entanglement. When the system and environment become incoherent, entanglement vanishes completely, ensuring coherence as a prerequisite for entanglement. This decoherence allows the appearance of entanglement between parts of the system and environments leaving the quantum procedure inefficient. In the case of the measures, all reflected the same qualitative dynamical map, suggesting consistency and cogency in the findings.\\
In summary, our results show that the entanglement and coherence must be defined and controlled for quantum technology to be useful. They not only serve as a resource for processing quantum information, but proper knowledge of them can also lead to a better understanding of the related terms. This includes quantum states, surrounding environments, system-environment interactions, memory properties, quantum system compatibility with specific protocols, the interrelation between different quantum phenomena and, most significantly, engineering the procedures to achieve the required results.
\section{Appendix}
In this section, we give the details of the final density matrices for the four qubit $GHZ$ state when coupled to different classical environmental configurations driven by ORU noise. By using the Eq.\eqref{final density matrix of CSE}, the corresponding final density matrix for the $CSE$ configuration takes the form as:
\begin{align}
\rho_{CSE}(\tau)=\left[
\begin{array}{cccccccccccccccc}
 \mathcal{X}_1 & 0 & 0 & \mathcal{X}_2 & 0 & \mathcal{X}_2 & \mathcal{X}_2 & 0 & 0 & \mathcal{X}_2 & \mathcal{X}_2 & 0 & \mathcal{X}_2 & 0 & 0 & \mathcal{X}_1 \\
 0 & \mathcal{X}_3 & \mathcal{X}_3 & 0 & \mathcal{X}_3 & 0 & 0 & \mathcal{X}_3 & \mathcal{X}_3 & 0 & 0 & \mathcal{X}_3 & 0 & \mathcal{X}_3 & \mathcal{X}_3 & 0 \\
 0 & \mathcal{X}_3 & \mathcal{X}_3 & 0 & \mathcal{X}_3 & 0 & 0 & \mathcal{X}_3 & \mathcal{X}_3 & 0 & 0 & \mathcal{X}_3 & 0 & \mathcal{X}_3 & \mathcal{X}_3 & 0 \\
 \mathcal{X}_2 & 0 & 0 & \mathcal{X}_4 & 0 & \mathcal{X}_4 & \mathcal{X}_4 & 0 & 0 & \mathcal{X}_4 & \mathcal{X}_4 & 0 & \mathcal{X}_4 & 0 & 0 & \mathcal{X}_2 \\
 0 & \mathcal{X}_3 & \mathcal{X}_3 & 0 & \mathcal{X}_3 & 0 & 0 & \mathcal{X}_3 & \mathcal{X}_3 & 0 & 0 & \mathcal{X}_3 & 0 & \mathcal{X}_3 & \mathcal{X}_3 & 0 \\
 \mathcal{X}_2 & 0 & 0 & \mathcal{X}_4 & 0 & \mathcal{X}_4 & \mathcal{X}_4 & 0 & 0 & \mathcal{X}_4 & \mathcal{X}_4 & 0 & \mathcal{X}_4 & 0 & 0 & \mathcal{X}_2 \\
 \mathcal{X}_2 & 0 & 0 & \mathcal{X}_4 & 0 & \mathcal{X}_4 & \mathcal{X}_4 & 0 & 0 & \mathcal{X}_4 & \mathcal{X}_4 & 0 & \mathcal{X}_4 & 0 & 0 & \mathcal{X}_2 \\
 0 & \mathcal{X}_3 & \mathcal{X}_3 & 0 & \mathcal{X}_3 & 0 & 0 & \mathcal{X}_3 & \mathcal{X}_3 & 0 & 0 & \mathcal{X}_3 & 0 & \mathcal{X}_3 & \mathcal{X}_3 & 0 \\
 0 & \mathcal{X}_3 & \mathcal{X}_3 & 0 & \mathcal{X}_3 & 0 & 0 & \mathcal{X}_3 & \mathcal{X}_3 & 0 & 0 & \mathcal{X}_3 & 0 & \mathcal{X}_3 & \mathcal{X}_3 & 0 \\
 \mathcal{X}_2 & 0 & 0 & \mathcal{X}_4 & 0 & \mathcal{X}_4 & \mathcal{X}_4 & 0 & 0 & \mathcal{X}_4 & \mathcal{X}_4 & 0 & \mathcal{X}_4 & 0 & 0 & \mathcal{X}_2 \\
 \mathcal{X}_2 & 0 & 0 & \mathcal{X}_4 & 0 & \mathcal{X}_4 & \mathcal{X}_4 & 0 & 0 & \mathcal{X}_4 & \mathcal{X}_4 & 0 & \mathcal{X}_4 & 0 & 0 & \mathcal{X}_2 \\
 0 & \mathcal{X}_3 & \mathcal{X}_3 & 0 & \mathcal{X}_3 & 0 & 0 & \mathcal{X}_3 & \mathcal{X}_3 & 0 & 0 & \mathcal{X}_3 & 0 & \mathcal{X}_3 & \mathcal{X}_3 & 0 \\
 \mathcal{X}_2 & 0 & 0 & \mathcal{X}_4 & 0 & \mathcal{X}_4 & \mathcal{X}_4 & 0 & 0 & \mathcal{X}_4 & \mathcal{X}_4 & 0 & \mathcal{X}_4 & 0 & 0 & \mathcal{X}_2 \\
 0 & \mathcal{X}_3 & \mathcal{X}_3 & 0 & \mathcal{X}_3 & 0 & 0 & \mathcal{X}_3 & \mathcal{X}_3 & 0 & 0 & \mathcal{X}_3 & 0 & \mathcal{X}_3 & \mathcal{X}_3 & 0 \\
 0 & \mathcal{X}_3 & \mathcal{X}_3 & 0 & \mathcal{X}_3 & 0 & 0 & \mathcal{X}_3 & \mathcal{X}_3 & 0 & 0 & \mathcal{X}_3 & 0 & \mathcal{X}_3 & \mathcal{X}_3 & 0 \\
 \mathcal{X}_1 & 0 & 0 & \mathcal{X}_2 & 0 & \mathcal{X}_2 & \mathcal{X}_2 & 0 & 0 & \mathcal{X}_2 & \mathcal{X}_2 & 0 & \mathcal{X}_2 & 0 & 0 & \mathcal{X}_1
\end{array}
\right],
\end{align}
where
\begin{align*}
\mathcal{X}_1=&\frac{19}{64}+\eta_1+\eta_2, &  \mathcal{X}_2=&-\frac{1}{64}(1+\eta_3)+\eta_4,\\
\mathcal{X}_3=&\frac{1}{64}(1-\eta_5), & \mathcal{X}_4=&\frac{1}{64}(1+\eta_6)-\eta_7.\\
\end{align*}
By utilizing the Eq.\eqref{final density matrix of BSE}, final density matrix of the $BSE$ configuration is given as:
\begin{align}
\rho_{BSE}(\tau)=\left[
\begin{array}{cccccccccccccccc}
 \mathcal{P}_1 & 0 & 0 & \mathcal{P}_2 & 0 & 0 & 0 & 0 & 0 & 0 & 0 & 0 & \mathcal{P}_2 & 0 & 0 & \mathcal{P}_1 \\
 0 & \mathcal{P}_3 & \mathcal{P}_3 & 0 & 0 & 0 & 0 & 0 & 0 & 0 & 0 & 0 & 0 & \mathcal{P}_3 & \mathcal{P}_3 & 0 \\
 0 & \mathcal{P}_3 & \mathcal{P}_3 & 0 & 0 & 0 & 0 & 0 & 0 & 0 & 0 & 0 & 0 & \mathcal{P}_3 & \mathcal{P}_3 & 0 \\
 \mathcal{P}_2 & 0 & 0 & \mathcal{P}_4 & 0 & 0 & 0 & 0 & 0 & 0 & 0 & 0 & \mathcal{P}_4 & 0 & 0 & \mathcal{P}_2 \\
 0 & 0 & 0 & 0 & \mathcal{P}_5 & 0 & 0 & \mathcal{P}_5 & \mathcal{P}_5 & 0 & 0 & \mathcal{P}_5 & 0 & 0 & 0 & 0 \\
 0 & 0 & 0 & 0 & 0 & \mathcal{P}_6 & \mathcal{P}_6 & 0 & 0 & \mathcal{P}_6 & \mathcal{P}_6 & 0 & 0 & 0 & 0 & 0 \\
 0 & 0 & 0 & 0 & 0 & \mathcal{P}_6 & \mathcal{P}_6 & 0 & 0 & \mathcal{P}_6 & \mathcal{P}_6 & 0 & 0 & 0 & 0 & 0 \\
 0 & 0 & 0 & 0 & \mathcal{P}_5 & 0 & 0 & \mathcal{P}_5 & \mathcal{P}_5 & 0 & 0 & \mathcal{P}_5 & 0 & 0 & 0 & 0 \\
 0 & 0 & 0 & 0 & \mathcal{P}_5 & 0 & 0 & \mathcal{P}_5 & \mathcal{P}_5 & 0 & 0 & \mathcal{P}_5 & 0 & 0 & 0 & 0 \\
 0 & 0 & 0 & 0 & 0 & \mathcal{P}_6 & \mathcal{P}_6 & 0 & 0 & \mathcal{P}_6 & \mathcal{P}_6 & 0 & 0 & 0 & 0 & 0 \\
 0 & 0 & 0 & 0 & 0 & \mathcal{P}_6 & \mathcal{P}_6 & 0 & 0 & \mathcal{P}_6 & \mathcal{P}_6 & 0 & 0 & 0 & 0 & 0 \\
 0 & 0 & 0 & 0 & \mathcal{P}_5 & 0 & 0 & \mathcal{P}_5 & \mathcal{P}_5 & 0 & 0 & \mathcal{P}_5 & 0 & 0 & 0 & 0 \\
 \mathcal{P}_2 & 0 & 0 & \mathcal{P}_4 & 0 & 0 & 0 & 0 & 0 & 0 & 0 & 0 & \mathcal{P}_4 & 0 & 0 & \mathcal{P}_2 \\
 0 & \mathcal{P}_3 & \mathcal{P}_3 & 0 & 0 & 0 & 0 & 0 & 0 & 0 & 0 & 0 & 0 & \mathcal{P}_3 & \mathcal{P}_3 & 0 \\
 0 & \mathcal{P}_3 & \mathcal{P}_3 & 0 & 0 & 0 & 0 & 0 & 0 & 0 & 0 & 0 & 0 & \mathcal{P}_3 & \mathcal{P}_3 & 0 \\
 \mathcal{P}_1 & 0 & 0 & \mathcal{P}_2 & 0 & 0 & 0 & 0 & 0 & 0 & 0 & 0 & \mathcal{P}_2 & 0 & 0 & \mathcal{P}_1
\end{array}
\right]
\end{align}
where
\begin{align*}
\mathcal{P}_1=&\frac{1}{32} \kappa_1 \left(1+\kappa_2)\right), & \mathcal{P}_2=&\frac{1}{32} \kappa_3 \left(\kappa_4\right),\\
\mathcal{P}_3=&\frac{1}{32}(-1+ \kappa_5\left(\kappa_6\right)),& \mathcal{P}_4=&\frac{1}{32} \kappa_7 \left(1+\kappa_8\right),\\
\mathcal{P}_5=&\frac{1}{32} \kappa_9 \left(-1+\kappa_{10} \text{Sinh}[\beta ]\right).
\end{align*}
For the $TSE$ configuration, we can get the final density matrix by using the Eq.\eqref{final density matrix of TSE} as:
\begin{align}
\rho_{TSE}(\tau)=\left[
\begin{array}{cccccccccccccccc}
 \mathcal{Q}_1 & 0 & 0 & \mathcal{Q}_2 & 0 & \mathcal{Q}_2 & \mathcal{Q}_2 & 0 & 0 & \mathcal{Q}_2 & \mathcal{Q}_2 & 0 & \mathcal{Q}_2 & 0 & 0 & \mathcal{Q}_1 \\
 0 & \mathcal{Q}_3 & \mathcal{Q}_2 & 0 & \mathcal{Q}_2 & 0 & 0 & \mathcal{Q}_2 & \mathcal{Q}_2 & 0 & 0 & \mathcal{Q}_2 & 0 & \mathcal{Q}_2 & \mathcal{Q}_3 & 0 \\
 0 & \mathcal{Q}_2 & \mathcal{Q}_4 & 0 & \mathcal{Q}_4 & 0 & 0 & \mathcal{Q}_4 & \mathcal{Q}_4 & 0 & 0 & \mathcal{Q}_4 & 0 & \mathcal{Q}_4 & \mathcal{Q}_2 & 0 \\
 \mathcal{Q}_2 & 0 & 0 & \mathcal{Q}_4 & 0 & \mathcal{Q}_4 & \mathcal{Q}_4 & 0 & 0 & \mathcal{Q}_4 & \mathcal{Q}_4 & 0 & \mathcal{Q}_4 & 0 & 0 & \mathcal{Q}_2 \\
 0 & \mathcal{Q}_2 & \mathcal{Q}_4 & 0 & \mathcal{Q}_4 & 0 & 0 & \mathcal{Q}_4 & \mathcal{Q}_4 & 0 & 0 & \mathcal{Q}_4 & 0 & \mathcal{Q}_4 & \mathcal{Q}_2 & 0 \\
 \mathcal{Q}_2 & 0 & 0 & \mathcal{Q}_4 & 0 & \mathcal{Q}_4 & \mathcal{Q}_4 & 0 & 0 & \mathcal{Q}_4 & \mathcal{Q}_4 & 0 & \mathcal{Q}_4 & 0 & 0 & \mathcal{Q}_2 \\
 \mathcal{Q}_2 & 0 & 0 & \mathcal{Q}_4 & 0 & \mathcal{Q}_4 & \mathcal{Q}_4 & 0 & 0 & \mathcal{Q}_4 & \mathcal{Q}_4 & 0 & \mathcal{Q}_4 & 0 & 0 & \mathcal{Q}_2 \\
 0 & \mathcal{Q}_2 & \mathcal{Q}_4 & 0 & \mathcal{Q}_4 & 0 & 0 & \mathcal{Q}_4 & \mathcal{Q}_4 & 0 & 0 & \mathcal{Q}_4 & 0 & \mathcal{Q}_4 & \mathcal{Q}_2 & 0 \\
 0 & \mathcal{Q}_2 & \mathcal{Q}_4 & 0 & \mathcal{Q}_4 & 0 & 0 & \mathcal{Q}_4 & \mathcal{Q}_4 & 0 & 0 & \mathcal{Q}_4 & 0 & \mathcal{Q}_4 & \mathcal{Q}_2 & 0 \\
 \mathcal{Q}_2 & 0 & 0 & \mathcal{Q}_4 & 0 & \mathcal{Q}_4 & \mathcal{Q}_4 & 0 & 0 & \mathcal{Q}_4 & \mathcal{Q}_4 & 0 & \mathcal{Q}_4 & 0 & 0 & \mathcal{Q}_2 \\
 \mathcal{Q}_2 & 0 & 0 & \mathcal{Q}_4 & 0 & \mathcal{Q}_4 & \mathcal{Q}_4 & 0 & 0 & \mathcal{Q}_4 & \mathcal{Q}_4 & 0 & \mathcal{Q}_4 & 0 & 0 & \mathcal{Q}_2 \\
 0 & \mathcal{Q}_2 & \mathcal{Q}_4 & 0 & \mathcal{Q}_4 & 0 & 0 & \mathcal{Q}_4 & \mathcal{Q}_4 & 0 & 0 & \mathcal{Q}_4 & 0 & \mathcal{Q}_4 & \mathcal{Q}_2 & 0 \\
 \mathcal{Q}_2 & 0 & 0 & \mathcal{Q}_4 & 0 & \mathcal{Q}_4 & \mathcal{Q}_4 & 0 & 0 & \mathcal{Q}_4 & \mathcal{Q}_4 & 0 & \mathcal{Q}_4 & 0 & 0 & \mathcal{Q}_2 \\
 0 & \mathcal{Q}_2 & \mathcal{Q}_4 & 0 & \mathcal{Q}_4 & 0 & 0 & \mathcal{Q}_4 & \mathcal{Q}_4 & 0 & 0 & \mathcal{Q}_4 & 0 & \mathcal{Q}_4 & \mathcal{Q}_2 & 0 \\
 0 & \mathcal{Q}_3 & \mathcal{Q}_2 & 0 & \mathcal{Q}_2 & 0 & 0 & \mathcal{Q}_2 & \mathcal{Q}_2 & 0 & 0 & \mathcal{Q}_2 & 0 & \mathcal{Q}_2 & \mathcal{Q}_3 & 0 \\
 \mathcal{Q}_1 & 0 & 0 & \mathcal{Q}_2 & 0 & \mathcal{Q}_2 & \mathcal{Q}_2 & 0 & 0 & \mathcal{Q}_2 & \mathcal{Q}_2 & 0 & \mathcal{Q}_2 & 0 & 0 & \mathcal{Q}_1
\end{array}
\right]
\end{align}
where \begin{align*}
\mathcal{Q}_1=&\frac{1}{32} \left(5+\lambda_1\right), & \mathcal{Q}_2=&\frac{1}{32} \left(-1+\lambda_2\right),\\
\mathcal{Q}_3=&\frac{1}{32} \left(5-\lambda_3\right), & \mathcal{Q}_4=&\frac{1}{16} \lambda_4\sinh[\beta ].
\end{align*}
Finally, we can obtain the final density matrix for the $TSE$ configuration has the form as:
\begin{align}
\rho_{ISE}(\tau)=\left[
\begin{array}{cccccccccccccccc}
 \mathcal{R}_1 & 0 & 0 & 0 & 0 & 0 & 0 & 0 & 0 & 0 & 0 & 0 & 0 & 0 & 0 & \mathcal{R}_1 \\
 0 & \mathcal{R}_2 & 0 & 0 & 0 & 0 & 0 & 0 & 0 & 0 & 0 & 0 & 0 & 0 & \mathcal{R}_2 & 0 \\
 0 & 0 & \mathcal{R}_2 & 0 & 0 & 0 & 0 & 0 & 0 & 0 & 0 & 0 & 0 & \mathcal{R}_2 & 0 & 0 \\
 0 & 0 & 0 & \mathcal{R}_3 & 0 & 0 & 0 & 0 & 0 & 0 & 0 & 0 & \mathcal{R}_3 & 0 & 0 & 0 \\
 0 & 0 & 0 & 0 & \mathcal{R}_2 & 0 & 0 & 0 & 0 & 0 & 0 & \mathcal{R}_2 & 0 & 0 & 0 & 0 \\
 0 & 0 & 0 & 0 & 0 & \mathcal{R}_3 & 0 & 0 & 0 & 0 & \mathcal{R}_3 & 0 & 0 & 0 & 0 & 0 \\
 0 & 0 & 0 & 0 & 0 & 0 & \mathcal{R}_3 & 0 & 0 & \mathcal{R}_3 & 0 & 0 & 0 & 0 & 0 & 0 \\
 0 & 0 & 0 & 0 & 0 & 0 & 0 & \mathcal{R}_2 & \mathcal{R}_2 & 0 & 0 & 0 & 0 & 0 & 0 & 0 \\
 0 & 0 & 0 & 0 & 0 & 0 & 0 & \mathcal{R}_2 & \mathcal{R}_2 & 0 & 0 & 0 & 0 & 0 & 0 & 0 \\
 0 & 0 & 0 & 0 & 0 & 0 & \mathcal{R}_3 & 0 & 0 & \mathcal{R}_3 & 0 & 0 & 0 & 0 & 0 & 0 \\
 0 & 0 & 0 & 0 & 0 & \mathcal{R}_3 & 0 & 0 & 0 & 0 & \mathcal{R}_3 & 0 & 0 & 0 & 0 & 0 \\
 0 & 0 & 0 & 0 & \mathcal{R}_2 & 0 & 0 & 0 & 0 & 0 & 0 & \mathcal{R}_2 & 0 & 0 & 0 & 0 \\
 0 & 0 & 0 & \mathcal{R}_3 & 0 & 0 & 0 & 0 & 0 & 0 & 0 & 0 & \mathcal{R}_3 & 0 & 0 & 0 \\
 0 & 0 & \mathcal{R}_2 & 0 & 0 & 0 & 0 & 0 & 0 & 0 & 0 & 0 & 0 & \mathcal{R}_2 & 0 & 0 \\
 0 & \mathcal{R}_2 & 0 & 0 & 0 & 0 & 0 & 0 & 0 & 0 & 0 & 0 & 0 & 0 & \mathcal{R}_2 & 0 \\
 \mathcal{R}_1 & 0 & 0 & 0 & 0 & 0 & 0 & 0 & 0 & 0 & 0 & 0 & 0 & 0 & 0 & \mathcal{R}_1
\end{array}
\right]
\end{align}
where
\begin{align*}
\mathcal{R}_1=&\frac{1}{16} \varsigma_1 \left(1+ \varsigma_2 \right), & \mathcal{R}_2=&\frac{1}{16} \left(1-\varsigma_3 \right),\\
\mathcal{R}_3=&\frac{1}{16} \varsigma_4 \left(-1+\varsigma_5 \right)^2.\\
\end{align*}

\end{document}